\documentclass[a4paper]{ESLAB2005}
\usepackage{epsfig}

\begin{document}
 
\title{Trends in Space Astronomy and Cosmic Vision 2015-2025}

\author{C. Turon\inst{1}, C. Done\inst{2}, A. Quirrenbach\inst{3}, P. Schneider\inst{4}, C. Aerts\inst{5}, A. Bazzano\inst{6}, \\
J. Cernicharo\inst{7},  P. de Bernardis\inst{8}, A. Goobar\inst{9},  T. Henning\inst{10}, R.\,J. Ivison\inst{11},  J.-P. Kneib\inst{12}, E. Meurs\inst{13}, \\
M. van der Klis\inst{14}, P. Viana\inst{15} S. Volont\`{e}\inst{16}  \and W.\,W. Zeilinger\inst{17}} 
  \institute{GEPI - UMR CNRS 8111, Observatoire de Paris, Section  de Meudon, 92195 MEUDON cedex, France
  \and Department of Physics, University of Durham, South Road, DURHAM,  DH1 3LE, UK
  \and Sterrewacht Leiden, Postbus 9513, 2300 RA  LEIDEN, The Netherlands
  \and Institut f\"{u}r Astrophysik und extraterrestrische Forschung, Universit\"{a}t Bonn, Auf dem H\"{u}gel 71, 53121 BONN
  \and Faculteit Wetenschappen, Instituut voor Sterrenkunde, Celestijnenlaan 200B, 3001 LEUVEN, Belgium
  \and Istituto di Astrofisica Spaziale e Fisica Cosmica, Area di Ricerca di Roma 2, Tor Vergata, Via del Fosso del Cavaliere 100, 00133 ROMA, Italy
  \and CSIC. IEM, Dpt. Molecular and Infrared Astrophysics, C/Serrano 121, 28006 MADRID, Spain
  \and Dipartimento di Fisica, Universit\`{a} La Sapienza, P.le A. Moro 2, 00185 ROMA, Italy
  \and SCFAB - Department of Physics, Stockholm University, Roslagstullsbacken 21, 106 91 STOCKHOLM, Sweden
  \and Max-Planck-Institut f\"{u}r Astronomie, K\"{o}nigstuhl 17, 69117 HEIDELBERG, Germany
  \and UK Astronomy Technology Centre, Royal Observatory, Blackford Hill, EDINBURGH EH9 3HJ, UK
  \and Laboratoire dÕAstrophysique de Marseille, Traverse du Siphon Ð B.P. 8, 13376 MARSEILLE cedex 12, France
  \and Dublin Institute for Advanced Studies, School of Cosmic Physics, Dunsink Observatory, DUBLIN 15, Ireland
  \and Astronomical Institute Anton Pannekoek, University of Amsterdam, Kruislaan 403, 1908 SJ AMSTERDAM, The Netherlands
  \and Centro de Astrofisica-Universidade do Porto, Rua das Estrelas, 4150-762 PORTO, Portugal
  \and European Space Agency, 8-10 rue Mario-Nikis, 75738 PARIS cedex 15, France
  \and Institut f\"{u}r Astronomie der Universit\"{a}t Wien, T\"{u}rkenschanzstrasse 17, 1180 WIEN, Austria}

\maketitle 

\begin{abstract}
As a short introduction to the astronomy session, the response of the community to the Call for Themes issued by ESA and the specific themes selected by the Astronomy Working Group are briefly presented in connection with the four grand themes finally selected for the ESA Science Programme. They are placed in the context of the main discoveries of the past decade and the astronomy projects currently in their development or definition phase. Finally, possible strategies for their implementation are summarised.

\keywords{Stars: formation -- exoplanets: formation -- exoplanets: detection -- exoplanets: characterisation -- early Universe  -- dark matter -- cosmic microwave background -- large-scale structure of Universe -- Black holes -- Galaxies: evolution -- supernovae}
\end{abstract}

\section{Introduction}
\label{sec:intro}
 In April 2004, the ESA Science Directorate initiated its long term planning cycle for the next decade referred to as Cosmic Vision 2015-2025.  The process was started with a Call for Scientific Themes of the future ESA Science Programme released on 27 April 2004. The response of the scientific community consisted in 151 proposals of which 47 dealt with astronomy topics. This greatly exceeded the corresponding numbers from previous calls (Horizon 2000, \cite{bb84}; Horizon 2000-Plus, \cite{bw94}). The Astronomy Working Group (AWG), in charge of evaluating the astronomy proposals, was overwhelmed by the quality and the quantity of the strong community response. This provided unambiguous evidence for the era of great discoveries that astronomy is currently undergoing. The responses covered a large range of topics, from the physics of nearby stars and exo-planetary systems to the Universe as a whole in all wavelength regimes of the electromagnetic spectrum. 
 
The proposals were evaluated taking into account the current and near-future instrumental development in astronomy, with particular regards to the ESA missions HST, XMM-Newton, Integral, Herschel, Planck, GAIA and JWST, as well as  the great scientific breakthroughs witnessed in recent times or expected to be achieved with ongoing or planned missions over the next decade. The AWG identified three major Scientific Themes believed to be at the forefront of astrophysical research for the period 2015-2025 and for which space missions will be mandatory in order to achieve the huge steps forward required for their understanding.

The AWG was well aware of the fact that it was not possible to include all strong and original ideas in the three major Themes. However, the AWG believed that some issues missing could be revisited in the future as new scientific opportunities would arise.

\section{Three major Themes in Astronomy}
\label{sec:themes}
The proposals were evaluated on the basis of essential questions such as: What is new? What is the likely impact in the domain?  What is the likely impact on science? What is the expected range of application? What is the added value of space? In addition, the readiness of the required technologies was considered in order to propose possible strategies and distinguish between short-term (around 2015), medium--term (2020), long-term (2025) and very long-term projects ($>$ 2025).

The three main themes identified by the Astronomy Working Group are:
\begin{enumerate}
  \item Other worlds and life in the Universe. Placing the Solar System into context
    \begin{itemize}
       \item Detection, census and characterisation of exoplanets
       \item Search for extraterrestrial life
       \item Formation of stars and planetary systems
    \end{itemize}   
  \item The Early Universe
    \begin{itemize}
      \item Investigating Dark Energy
      \item Probing inflation
      \item Observing the Universe taking shape
    \end{itemize}
  \item The evolving violent Universe
   \begin{itemize}
      \item Matter under extreme conditions
      \item Black holes and galaxy evolution
      \item Supernovae and the life cycle of matter
    \end{itemize}
\end{enumerate}

In looking for synergies and coherence, these themes were merged with themes from other disciplines to form the four themes selected for the new Cosmic Vision plan of the ESA Science Programme, namely: What are the conditions for planetary formation and the emergence of life? How does the Solar System work? What are the fundamental physical laws of the Universe? How did the Universe originate and what is it made of?

\section{Planetary formation and the emergence of life}
\label{sec:formation}
A decade ago, when the Solar System was the only planetary system known, theories were developed to account for the formation and evolution of such a system. Since then, the discovery of over 160 planets orbiting stars other than the Sun has taught us the limit of such an approach! The formation of many of these systems, with giant planets~-  `hot Jupiters' - orbiting close to the parent stars, seemed simply impossible within the framework of the theories accepted as little as ten years ago. Based on this salutary example, we can only wonder what scientific and philosophical revolution the discovery of life on another planet will provoke. For the first time since the dawn of philosophical and scientific thought, it is within our grasp to place the Solar System into the overall context of planetary formation, aiming at comparative planetology, and answer, rigourously and quantitatively, two fundamental questions: Are there other forms of life in the Solar System and did they have an independent origin from those that developed on Earth? Are there other planets orbiting other stars similar to our own Earth, and could they harbour life? The second of these questions is in the astronomy domain, with two main aspects: the detection of exoplanets and biomarkers, and the understanding of the star and planet formation process. 

\subsection{From exoplanets to biomarkers}
\label{subsec:exoplanets}
After the first discovery of an exoplanet in 1995, there has been steady progress towards detecting planets with ever- smaller masses, and towards the development of a broader suite of techniques to characterise their properties. All the discoveries of planets have so far come from ground-based telescopes, although space-based instrumentation has already provided some extraordinary insights such as Hubble observations of a photometric transit of one exoplanet in front of its mother star, and the evaporation of the atmosphere of another exoplanet. The situation is about to change with the prospective detection of planets of nearly the same size as the Earth by the French-ESA Corot mission and later by NASA's Kepler. Then GAIA will deliver important insights into the abundance of giant planets; the existence and location of such planets is crucial for the possible existence of Earth-like planets in the habitable zone. GAIA will also further improve our understanding of the stellar and Galactic constraints on planet formation and existence.

The next major break-through in exoplanetary science will be the detection and detailed characterisation of Earth-like planets in habitable zones. The prime goal would be to detect light from Earth-like planets and to perform low-resolution spectroscopy of their atmospheres in order to characterise their physical and chemical properties. 
Only a space observatory will have the ability to distinguish the light from Earth-like planets and to perform the low-resolution spectroscopy of their atmospheres needed to characterise their physical and chemical properties. The target sample would include about 200 stars in the solar neighbourhood. Follow-up spectroscopy covering the molecular bands of CO$_2$, H$_2$O, O$_3$, and CH$_4$, typical tracers of the Earth spectrum, will deepen our understanding of Earth-like planets in general, and may lead to the identification of unique biomarkers. The search for life on other planets will enable us to place life as it exists today on Earth in the context of planetary and biological evolution and survival. 

A near infrared nulling interferometer operating in the wavelength range of 6 to 20 microns would provide the tool necessary to achieve these objectives. Based on the technology and expertise already being developed, and implemented around 2015, it would make Europe a pioneer in this field and guarantee its continuing leadership in exo\-planet research.

On a longer time-scale, a complete census of all Earth-sized planets within 100 parsecs of the Sun would be highly desirable. Building on GAIA's expected contribution on larger planets, this could be achieved with a high-precision terrestrial planet astrometric surveyor. Eventually the direct detection of such planets followed by high-resolution spectroscopy with a large telescope at infrared, visible and ultraviolet wavelengths, and ultimately by spatially resolved imaging, will mark the coming of age of yet another entirely new field of astronomy: comparative exo\-planetology.

\subsection{From gas and dust to stars and planets}
\label{subsec:stars}
The formation of planetary systems has to be understood in the wider context of the process of star formation and the evolution of proto-planetary disks. We presently lack a generally accepted scenario for the formation of solar-type stars, with magnetic fields and turbulence being discussed as possible key factors in the birth process. We also have to realise that the newly detected `planetary worlds' are quite different from our own Solar System. These new findings cast doubt on existing theories of the formation of planets and planetary systems, which have been based almost entirely on the single case of the Solar System. The insight gained from observations of exoplanets has demonstrated the importance of planet-disk and planet-planet interactions, leading to such surprising results as the inward migration of giant planets and the question of how stable such systems actually are. 

The understanding of the star and planet formation process requires a multi-wavelength approach, mostly from near-infrared to millimetre wavelengths. With the ISO mission, the upcoming Herschel observatory, the strong participation in JWST, and with the ground-based facilities of ESO and the ALMA project, the European star formation community is in a very strong scientific position. In order to make further progress in the field, and to complement the capabilities of ALMA, a far-IR mission is required. ALMA will not be able to access the important water lines, a key goal of a far-infrared mission. In addition, the peak of the spectral energy distribution is located at far-infrared wavelengths. Such a mission would need to provide spatial resolution of the order of 0.01 arcsec to resolve protostars and disks in the nearest star-forming regions, and should be able to perform high- and low-resolution spectroscopy to characterise line emission and dust mineralogy.

\section{The early and evolving violent Universe}
\label{sec:Universe}
As a consequence of a fantastic increase in our knowledge of the Universe in the past two decades, fundamental questions can at last now be better identified and formulated: detecting imprints of the very early stages of the evolution of the Universe in radiation observable today; unravelling the nature of dark energy and dark matter; understanding how the observable Universe took shape; and how it evolves through the violent mechanisms taking place in interaction with black holes and neutron stars.  These scientific issues will remain at the centre of cosmological and astrophysical interest for at least the next two decades. 

\subsection{The early Universe}
\label{subsec:early}
The totally unexpected discovery of a presently accelerating Universe leaves us with the quest for the driving force behind it. Termed `Dark Energy', this component of the Universe currently has no explanation in terms of a physical model and is presently the largest challenge for fundamental physics. Unravelling the nature of Dark Energy will most likely lead to uncovering new and deep laws of physics. The two best means of investigating Dark Energy as currently identified are the weak lensing effect caused by the large-scale matter distribution in the Universe and the study of the luminosity-redshift relation of Supernovae~Ia up to high redshifts. Precision measurements of both of these effects require a space-based wide-field optical and near-IR imaging capability.

At its origin, our Universe is believed to have undergone a phase of rapid acceleration, called inflation. Generic predictions of inflation, such as the spatial flatness and the shape of the initial density fluctuations, have impressively been confirmed with recent observations, most noticeably the WMAP results. However, the physical mechanism driving inflation is unclear at present and competing physical theories exist. Since they differ in their predictions about the amplitude and shape of the primordial gravitational wave spectrum, determining those provides the key for deciphering the very beginning of the cosmos. Observing the polarisation properties of the cosmic microwave background (CMB) is the best approach for characterising these primordial gravity waves within the 2015-2025 time frame. It requires a multi-frequency all-sky map of the CMB polarisation with much larger sensitivity than Planck will achieve, which, due to the weakness of the signal, makes space-based observations mandatory. 

On a longer time-scale, a post-LISA Gravitational Wave Cosmic Surveyor, operating in a new frequency window (0.1-1.0 Hz), with orders of magnitude more sensitivity than LISA, could make a key step toward the direct detection of the primordial gravitational wave background. These gravitational waves should be emitted at the end of the inflation era, and might even contain information about the Universe before inflation set in. 

\subsection{The Universe taking shape}
\label{subsec:shape}
The tracing of cosmic history back to the time when the first luminous sources ignited, thus ending the dark ages of the Universe, has just begun. At that epoch the intergalactic medium was reionized, while large-scale structures increased in complexity, leading to the emergence of galaxies and their super-massive black holes. The merging of galaxies, their star-formation history, their relationship to quasars, and their interactions with the intergalactic medium are all processes that we have started to analyse with NASA-ESA's Hubble Space Telescope, ESA's XMM-Newton and NASA's Chandra, and other telescopes observing at complementary wavelengths, back to a time when the Universe had only about 10 \% of its current age. 

Pushing this history back to still earlier times will be one of the great achievements of Hubble's successor, the NASA-ESA-Canada James Webb Space Telescope. The rapid evolution of this research area requires the flexibility provided by observatory-type missions, including ESA's Herschel, and by the ground-based observatory ALMA. But even taking into account the gains of the next ten years, several questions will be left unanswered. In particular, the James Webb Space Telescope will miss the first clusters of galaxies and the precursors of quasars, expected to have central black holes with a much lower mass and luminosity than those seen closer in the cosmos. These will be best observed in by X-rays. 

As observational cosmology necessitates a multi-wave\-length approach, no single observatory can provide the complete cosmic picture. However, a large aperture X-ray observatory will be an early priority. It will be able to trace clusters of galaxies back to their formation epoch, making possible the study of the early heating and chemical enrichment of the intracluster gas, their relation to black hole activity, and the assembly of the clusters' galaxy population. Such an observatory should also be able to detect and characterise the precursors of quasars and locate the mergers of super-massive black holes expected to be detected by LISA. These objectives will require high sensitivity, with a collecting area above 10 m$^{2}$, and a wide field of view covering at least 5 arcmin to allow for work on extended objects. A field of 15 arcmin would substantially increase the serendipitous science return and the survey potential for locating the most extreme objects in the high-redshift Universe. High spatial resolution (under 5 arcsec) will be needed to avoid source confusion. A soft X-ray spectroscopy capability should make possible the detection of the missing half of the baryons in the local Universe, most likely hidden in the warm-hot intergalactic medium. 

Although the James Webb Space Telescope will register the red-shifted visible light from very distant objects (redshifts up to z $\sim$ 10) it will miss the star-forming regions hidden by dust. They will be observable, in the longer term, only by a new-generation far infrared observatory. This instrument will be essential to resolve the far-infrared background glow into discrete sources and so locate as much as 50 \% of the star formation activity, which is currently hidden from our view by dust absorption. The far infrared observatory will also resolve star-formation regions in nearby galaxies, both isolated and interacting, and identify through spectroscopy the cooling of molecular clouds with primordial chemical composition. These goals call for a resolution of about 1.5 arcsec at wavelengths around 200 microns. 

Other interesting information, especially on the warm-hot intergalactic medium and supernovae of Type Ia at low redshifts, would be obtainable using high resolution ultraviolet spectroscopy.

\subsection{The evolving violent Universe}
\label{subsec:violent}
Nature offers astrophysicists the possibility of observing objects in much more extreme conditions, in terms of gravity, density and temperature, than anything feasible on Earth. On the one hand, black holes and neutron stars are unique laboratories where the laws of physics can be probed under these extreme conditions. On the other hand, the same objects were the driving engines of the birth and evolution of galaxies, of the creation of heavy elements such as iron, and more generally, of the transformation of the primordial hydrogen and helium from which stars and galaxies were first being formed. 

Recent results show that super-massive black holes exist in the cores of most galaxies and that there must be a direct link between the formation and evolution of the black holes and of their host galaxies. X-ray emission is produced as surrounding gas is accreted by the black hole. This high-energy radiation is not just a witness of the existence of the black holes but also probes the rate at which these black holes grow to their current huge masses. Systematic, high-sensitivity X-ray observations of these growing super-massive black holes along cosmic history will give unprecedented information on the growth of large-scale structures in the Universe and on the formation of galaxies. It is also of the utmost importance to understand the feedback between this process and the formation of stars and the galaxies themselves, for which the X-ray observations will need to be complemented by far infrared observations of the same objects to map star formation activity.

Thanks to another breakthrough of the past ten years, the extremely bright and brief emissions of gamma-ray bursts are now thought to be produced by a rapid accretion of gas onto newly formed black holes, resulting from the merging of neutron stars or the dramatic explosion of a high-mass supernova or hyper-nova. Capture of debris from the explosion results in an extreme rate of mass accretion, which can power an ultra-relativistic jet of matter. This process can be used to probe the formation rate of high-mass stars that give rise to the hyper-novae, out to very high redshifts and the epoch of galaxy formation. 

Debris escaping from the scene disperses the heavy elements formed by nucleosynthesis in the massive stars, into the interstellar and intergalactic medium. We can witness this process with full detail when it happens very close to us, i.e., in the remnants of Supernovae occurred in our own Galaxy. The transported energy also heats the gas and suppresses star formation. The chemical abundances in the gas on these large scales can be determined from X-ray line emission, and reflect the supernova rate integrated over time, while nuclear lines at gamma-ray energies from radioactive isotopes give a snapshot of recent activity. Comparison of the abundances in the local and high redshift Universe will show the evolution of chemical enrichment and the impact of supernova feedback of energy on the growth of large-scale structures in the Universe. Comparison of the abundances of elements in the gas of galaxies, clusters of galaxies and in the intergalactic medium will shed light on the life cycle of matter in the Universe. 

When two super-massive black holes merge in a galaxy, they produce X-rays and gravitational waves. Simultaneous observations of these events by the X-ray observatory and by the gravitational wave detector LISA would bring complementary information. By pinpointing the galaxy, the X-ray detection will resolve any uncertainties in direction in the gravitational wave signature, and establish the distance of the event unambiguously.

Most of the topics quoted above build on successes achieved with ESA's XMM-Newton. They are also being addressed by the USA (RXTE and Chandra) and Japanese (ASCA and ASTRO-E2) space observatories. For Europe to maintain the lead in understanding the physics of the violent Universe, the next major step requires a large aperture X-ray observatory of high sensitivity ($\sim$ 10 m$^{2}$ collecting area) over a broad bandpass, ideally 0.1-50 keV, in order to handle the large photon rates of a variety of events. High spatial resolution ($\sim$ 1-2~arcsec) will be needed to avoid source confusion, and time resolution down to a few microseconds to probe the relevant time-scales. These performances would, for example, allow to probe the abundances of clusters and groups of galaxies out to redshifts 1-2, and track changes in the accretion flows onto black holes. The specifications are compatible with those for a large aperture X-ray observatory applied to studies of the Universe taking shape.

Closer to us, the supernova history of our own Galaxy will soon be much clearer through the spectroscopic diagnostics of MeV lines detected by ESA's Integral mission. By the end of the 2015-25 period, or soon after, the next generation detectors at these high energies (bandpass 100-2000 keV) will have a sensitivity two orders of magnitude better than Integral's. They would enable a gamma-ray imaging observatory to complete the supernova history of the Milky Way Ð and then to do the same for all the galaxies in the Local Group. 

\section{Proposed strategies in astronomy}
\label{sec:strategies}
For each of the two main themes detailed above, the most powerful tools needed to achieve the progress envisioned have been identified. These tools may be observatory-type missions to address broad scientific issues which require long technical developments, or specialised missions to target specific scientific questions.

\subsection{Stars, planets and life}
In the coming decade, progress is expected in three main areas: the search for terrestrial planets with the French-ESA Corot mission and later by NASA's Kepler; the understanding of the star and planet formation process with ESA's Herschel and NASA-ESA's JWST in space and with ALMA on the ground; the statistical census of giant planets with ESA's Gaia and to lesser extent with NASA's SIM mission.

Implemented in the very first part of the 2015-2025 decade, a {\bf near-infrared nulling interferometer}, for the wavelength range between 6 and 20 micrometers, would make Europe highly competitive in the field. It should be able to identify Earth-like planets orbiting other stars, and perform low-resolution spectroscopy of their atmospheres in order to characterise their physical and chemical properties and detect critical biomarkers. 

In the timeframe after 2020, a {\bf far-infrared observa\-tory-type mission}, with a spatial resolution of the order of 0.01 arcsec would resolve protostars and disks in the nearest star-forming regions, and should be able to perform high- and low-resolution spectroscopy to characterise line emission and dust mineralogy.

For a longer term prospective, three other concepts would bring complementary knowledge. First, an extreme accuracy astrometry mission (a `super-Gaia') would make a systematic census of terrestrial planets in the Solar neighbourhood up to 100 pc and, in parallel, allow to go deep into the Galaxy archaeology, and drastically extend the base for distance scale determination. Next, high-resolution spectroscopy with a large telescope at infrared, visible and ultraviolet wavelengths would bring a detailed analysis of the atmospheres of terrestrial planets. And ultimately, a large optical interferometer, for example along the hyper-telescope concept, would spatially resolve and image exo-Earths, opening another entirely new field of astronomy: comparative exoplanetology.

\subsection{The early and evolving violent Universe}
Recent discoveries have transformed our view of the Universe: the observational confirmation of the existence of an early phase of accelerated expansion, called inflation; the totally unexpected discovery of the `Dark Energy' driving a new phase of acceleration of the Universe; the observation of galaxies at ever-increasing distances; the discovery of strong gravity effects around black holes. 

In the coming decade, progress is expected in the understanding of the initial density fluctuations with Planck; in observational cosmology with Hubble, XMM-Newton, and later with Herschel and JWST; and in probing accretion and ejection mechanisms taking place in black holes and neutron stars with XMM-Newton and Integral.

In the first years of the 2015-2025 decade, the prime task for a {\bf Large Aperture X-ray Observatory}, with high throughput, high angular, spectral and time resolution, will be able to detect and study thoroughly the imprints of the evolution of the Universe, especially those triggered by violent events. Two focussed missions could also be devoted to unravel the nature of Dark Energy, with an {\bf optical/near IR wide-field imager}, and to probe inflation with an all-sky multi-frequency {\bf CMB polarisation mapper}.

For the second part of the decade, a new generation {\bf far-IR observatory}, already noted above as a means of observing the birth of stars and planets in our own Galaxy, would also have a very important cosmological role in tracing the evolution of galaxies by resolving the far infrared background into discrete sources, and by revealing the star-formation activity hidden by dust absorption.

Finally, at the very end of the 2015-25 decade, a Gamma-ray Imaging Observatory, building on the experience of Integral, would support the detailed examination of black holes by the Large Aperture X-ray Observatory and the detailed understanding of the history of supernovae in our Galaxy and in the Local Group of galaxies.


\begin{thebibliography}{}

\bibitem[\protect\astroncite{Bonnet \& Bleeker}{1984}]{bb84}
Bonnet R.-M. \& Bleeker J., ESA-SP-1070, December 1984, N.~Longdon \& H. Olthof eds

\bibitem[\protect\astroncite{Bonnet \& Woltjer}{1994-95}]{bw94}
Bonnet R.-M. \& Woltjer L., ESA-SP-1180, November 1994, August 1995, B. Battrick ed

\end{thebibliography}
\end{document}